\documentclass[final,3p,times,twocolumn,fleqn]{elsarticle}

\usepackage{amssymb}

\usepackage{lineno}

\usepackage{amsmath}
\journal{Physics Letters B}

\begin{document}

\begin{frontmatter}



\title{Statistical combination of searches for the  $\rm{X^{\pm}(5568)}$ state decaying into $\rm{B^0_s \pi^\pm}$}



 \author[1]{Paolo Iengo \fnref{fn1}}
 \ead{paolo.iengo@cern.ch}
 \address[1]{CERN, Esplanade des Particules 1, 1211 Geneva 23, Switzerland}
 
\begin{abstract}

  A statistical combination of the search results for the narrow $\rm{X^{\pm}(5568)}$ resonance decaying into 
$\rm{B_s^0 \pi^+}$ is reported, based on published results from the ATLAS, CMS, CDF and LHCb Collaborations.
A structure in the invariant mass distribution of $\rm{B_s^0 \pi^+}$ has been observed by the D0 Collaboration with a mass value of 5568 MeV but 
not confirmed by any of the latest searches from the other Collaborations. 
Limits have been set on the 
relative production rate $\rm{\rho_X}$ of the $\rm{X^{\pm}(5568)}$ and $\rm{B^0_s}$ states, times the branching ratio for the decay 
$\rm{X^{\pm}(5568) \rightarrow B^0_s\pi^\pm}$. 

By applying a statistical combination of limits set by the three LHC experiments, we derive a limit at 95\% Confidence Level of $\rho_X < 0.0092$
for $p_T(B^0_s)>$ 10 GeV, and $\rho_X < 0.0091$ for $p_T(B^0_s)>$ 15 GeV, superseding the previous most stringent upper limits.
The effect of including the CDF result in the combination is also discussed.
\end{abstract}



\begin{keyword}
$\rm{X^{\pm}(5568)}$ \sep Exotic hadrons \sep Tetraquark 
\sep Exclusion limits \sep Statistical combination 



\end{keyword}

\end{frontmatter}


\section{Introduction}
\label{Intro}

The evidence, with a 5.1$\sigma$ significance, published by the D0 Collaboration in 2016~\cite{ref:D01}, of a charged narrow state X decaying to $\rm{B^0_s\pi^\pm}$, with a mass peaking at 5568 MeV/\rm{$c^2$} and a width of 21.9 MeV/\rm{$c^2$}, generated a large 
interest in the physics community studying exotic hadrons (see e.g.~\cite{ref:theory}), with a possible interpretation of the new resonance as a {\it{buds}} tetraquark state. 
If confirmed, this was the first tetraquark state with four different quarks.

The search for the same resonance by the LHCb Collaboration~\cite{ref:LHCb} did not show any evidence of a signal. Afterward, CDF published the result of a search for the $\rm{X^{\pm}(5568) \rightarrow B^0_s\pi^\pm }$ decay~\cite{ref:CDF} consistent with no signal evidence. More recently, CMS~\cite{ref:CMS} and ATLAS~\cite{ref:ATLAS} have performed similar analysis on larger datasets, both concluding with results in line with LHCb and CDF, with no statistically significant signal excess.
Finally, D0 increased the dataset, including the semileptonic decay channel for the $\rm{B^0_s}$~\cite{ref:D02}, confirming the results of the first analysis with a clear indication of a signal. 

Attempts to explain the D0 results have been proposed, interpreting the observed peak in the $\rm{B^0_s\pi^\pm}$ invariant mass distribution as coming from a kinematic reflection~\cite{ref:Xinterpretation0}. 

In absence of a signal, the three LHC experiments and CDF have set upper limits on the 
relative production rate of the $\rm{X^{\pm}(5568)}$ and $\rm{B^0_s}$ states, times the branching ratio for the decay $\rm{X^{\pm}(5568) \rightarrow B^0_s\pi^\pm}$, $\rm{\rho_X}$:

\begin{eqnarray}\nonumber
  \rho_X &=& \frac{\sigma(pp \rightarrow X + {\rm{anything}}) }{\sigma(pp \rightarrow B^0_s + {\rm{anything}})}
  \times \mathcal{B}(X \rightarrow B^0_s \pi^\pm) \\ \nonumber
&=& \frac{N(X)}{N(B^0_s)} \times \frac{1}{\epsilon_X}
\label{eq:eq1}
\end{eqnarray}

\noindent where $\rm{N(B^0_s)}$ and $\rm{N(X)}$ is the number of reconstructed $\rm{B^0_s}$ and $\rm{X^{\pm}(5568)}$ candidates, respectively, and $\epsilon_X$ is the relative efficiency for the reconstruction of $\rm{B^0_s}$ and X events.

In this work we perform a statistical combination of the existing upper limits on the parameter $\rm{\rho_X}$ from the three LHC experiments and, separately, from the LHC experiments and CDF.

\section{Review of experimental sarches for $\rm{X^{\pm}(5568)}$}
\label{Sec1}

The observation from D0 of the $\rm{X^{\pm}(5568)}$ resonance is based on 10.4 ${\rm{fb^{-1}}}$ at 1.96 TeV proton-antiproton collision data collected during the run II at Tevatron. $\rm{B^0_s}$ candidates are reconstructed either in the $\rm{B^0_s \rightarrow J/\psi \phi \rightarrow \mu^+ \mu^- K^+ K^-}$ channel (hadronic)~\cite{ref:D01} or in the $\rm{B^0_s \rightarrow D_s \mu^\pm Y \rightarrow \phi \pi^\pm \mu^\pm Y}$ channel (semileptonic, here Y indicates any other particle in the final state of the $\rm{B^0_s}$ decay)~\cite{ref:D02}, with the $\rm{\phi}$ decaying into a $\rm{K^+K^-}$ pair. A cut on the transverse momentum $\rm{p_T}$ of the $\rm{B^0_s \pi}$ pair: $\rm{p_T(B^0_s \pi^\pm) > }$ 10 GeV is applied together with a topological cut between the direction of the $\rm{B^0_s}$ and $\rm{\pi}$, called cone cut: $\rm{\Delta R(B^0_s \pi^\pm) = \sqrt{\Delta\eta^2 + \Delta\phi^2} < 0.3}$. For the semileptonic channel the cone cut is applied to the angle between the $\rm{\mu D_s}$ system and the $\rm{\pi^\pm}$.
In the second D0 publication~\cite{ref:D02} a detailed discussion on the effect of the cone cut selection is reported, concluding that,
although affecting the final number of selected candidates, it is not a crucial factor for the signal evidence.

From a fit to the invariant mass distribution of the selected $\rm{B^0_s \pi^\pm}$ pairs, $\rm{133 \pm 31}$ and $\rm{121^{+51}_{-34}}$ selected candidates are found for the hadronic and semileptonic analyses, respectively. 
Using a combined fit of the hadronic and semileptonic data, the measured mass of the state by D0 was
$\rm{5566.9^{+3.2}_{-3.1}(stat)^{+0.6}_{-1.2}(syst)}$ MeV/\rm{$c^2$} with a width of $\rm{\Gamma_X = 18.6^{+7.9}_{-6.1}(stat)^{+3.5}_{-3.8}(syst)}$ MeV/\rm{$c^2$}, 
and a significance of 6.7$\rm{\sigma}$.

LHCb searched for the $\rm{X^{\pm}(5568)}$ signal with the same two decay channels for the $\rm{B^0_s}$, in three bins of transverse momentum: $\rm{p_T(B^0_s)> 5, 10, 15}$ GeV and without applying a cone cut. The analysis was performed on 3 $\rm{fb^{-1}}$ at 7 and 8 TeV proton-proton collision data. Upper limits on $\rho_X$ from equation~\ref{eq:eq1} have been extracted with probability density function models for signal and background.

Similar analyses have been carried out by the CDF (9.6 $\rm{fb^{-1}}$ at 1.96 TeV proton-antiproton collision data), 
CMS (19.7 $\rm{fb^{-1}}$ at 8 TeV p-p collision data) and ATLAS 
(4.9 $\rm{fb^{-1}}$ at 7 TeV and 19.5 $\rm{fb^{-1}}$ at 8 TeV p-p collisions data) Collaborations, 
all looking at the $\rm{B^0_s \rightarrow J/\psi ~\phi }$ channel with $\rm{p_T(B^0_s)> 10}$ GeV for CDF and $\rm{p_T(B^0_s)> 10, 15}$ GeV for CMS and ATLAS. 
Similar to LHCb, no cone cut selection was applied in these analyses.

\begin{figure}[h!]
\centering
  \includegraphics[width=0.48\textwidth]{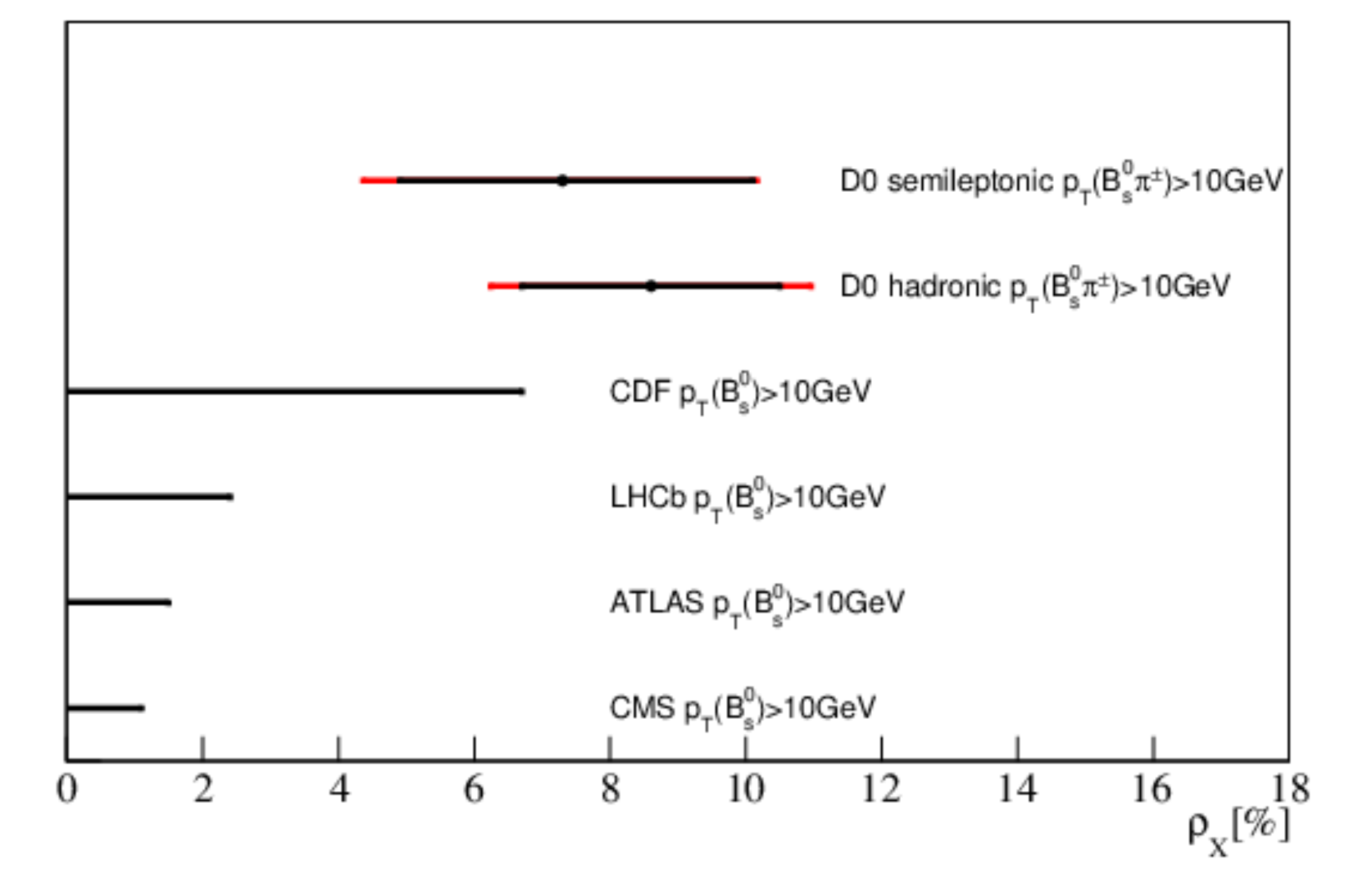}
  \includegraphics[width=0.48\textwidth]{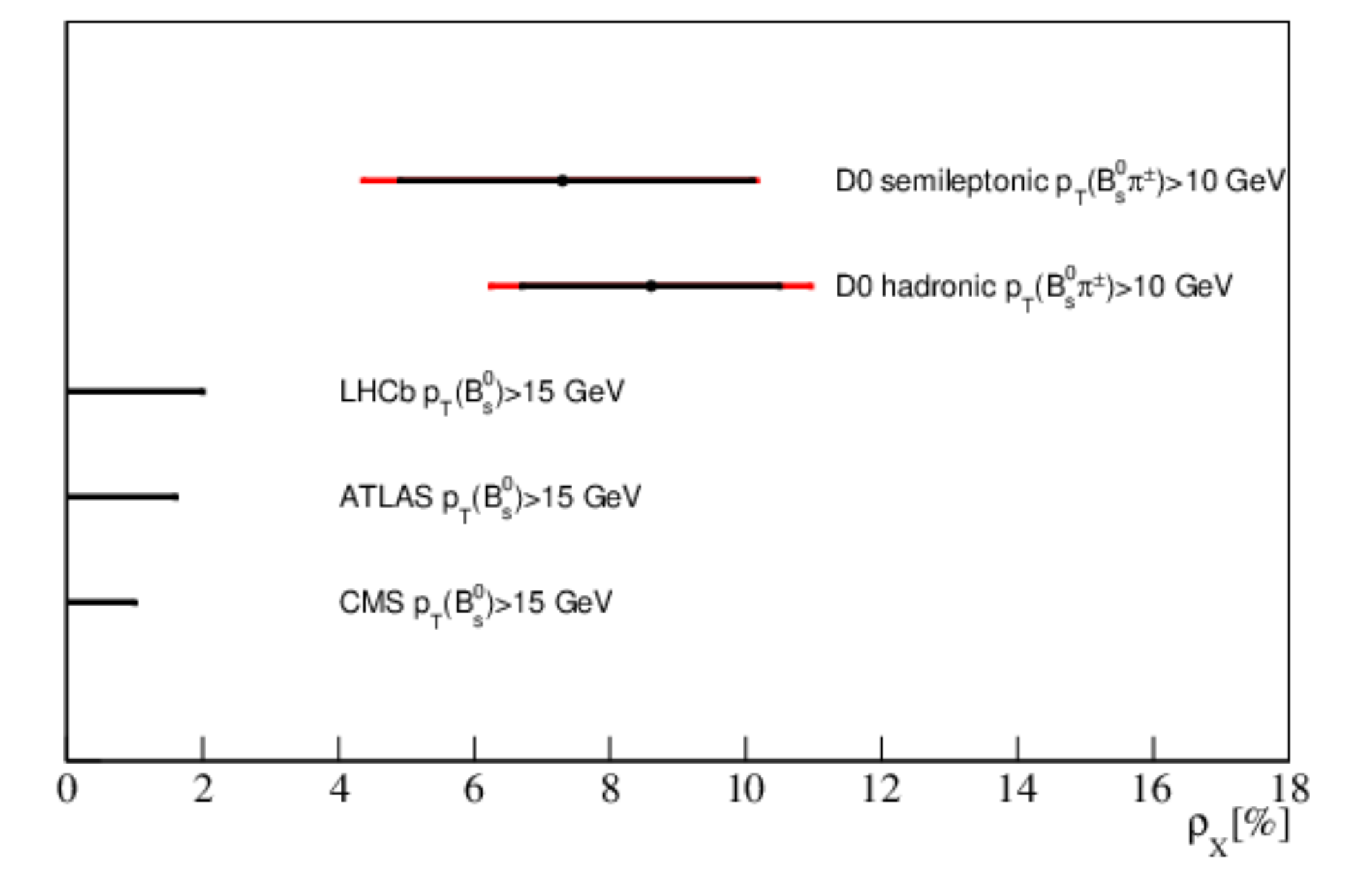}
\caption{Comparison of the measured values of $\rm{\rho_X}$ from D0 (with $\rm{p_T(B^0_s \pi)>}$ 10 GeV in both plots) and the upper limits at 95\% C.L. set by ATLAS, CDF, CMS and LHCb for $\rm{p_T(B^0_s)>}$ 10 GeV (top) and 15 GeV (bottom).
For the D0 measurements the black lines refer to statistical errors only, the red ones refer to statistical and systematic errors added in quadrature.  
}
\label{fig:figure1}
\end{figure}

Table~\ref{tab:table1} summarises for the five experiments, the number of reconstructed $\rm{B^0_s}$ and X(5568) candidates, 
and the relative efficiency $\rm{\epsilon_X}$ for each $\rm{B^0_s}$ decay channel and $\rm{p_T}$ cut. 
Figure~\ref{fig:figure1} shows the measured values (D0) and the 95\% Confidence Limits (LHCb, CDF, CMS, ATLAS) of the parameter $\rm{\rho_X}$.

\begin{table*}[h!!]
  \begin{center}
        \begin{tabular}{c|c|c|c|c|c}

\textbf{Experiment} & \textbf{$B^0_s$ decay channel} & \textbf{$p_T$ cut} & \textbf{$N(B^0_s)/10^3$} & \textbf{$N(X)$}  & \textbf{$\epsilon_X$} \\
\hline
\hline

D0 & $B^0_s \rightarrow D^-_s \pi^+$ & 10 GeV & $ 6222 \pm 141$ & $121^{+51}_{-34}$ & $0.266 \pm 0.040$ \\
D0 & $B^0_s \rightarrow J/\psi \phi$ & 10 GeV & $ 5582 \pm 100$ & $133 \pm 31$ & $0.277 \pm 0.040$ \\
\hline

LHCb & $B^0_s \rightarrow D^-_s \pi^+$ & 5 GeV & $62.2 \pm 0.3$ & $3 \pm 64$ & $0.127 \pm 0.002$ \\
LHCb & $B^0_s \rightarrow D^-_s \pi^+$ & 10 GeV & $28.4 \pm 0.2$ & $75 \pm 52$ & $0.213 \pm 0.003$ \\
LHCb & $B^0_s \rightarrow D^-_s \pi^+$ & 15 GeV & $8.8 \pm 0.1$ & $14 \pm 31$ & $0.289 \pm 0.005$ \\
\hline

LHCb & $B^0_s \rightarrow J/\psi \phi$ & 5 GeV & $46.3 \pm 0.2$ & $-33 \pm 43$ & $0.093 \pm 0.001$ \\
LHCb & $B^0_s \rightarrow J/\psi \phi$ & 10 GeV & $13.2 \pm 0.1$ & $12 \pm 33$ & $0.206 \pm 0.002$ \\
LHCb & $B^0_s \rightarrow J/\psi \phi$ & 15 GeV & $3.7 \pm 0.1$ & $-10 \pm 17$ & $0.290 \pm 0.004$ \\
\hline

CDF & $B^0_s \rightarrow J/\psi \phi$ & 10 GeV & $3.552 \pm 0.065$ & $36 \pm 33$ & $0.445 \pm 0.03$ \\
\hline

CMS & $B^0_s \rightarrow J/\psi \phi$ & 10 GeV & $49.277 \pm 0.278$ & $-85 \pm 160$ & $0.48 \pm 0.02$ \\
CMS & $B^0_s \rightarrow J/\psi \phi$ & 15 GeV & $40.292 \pm 0.246$ & $-103 \pm 230$ & $0.524 \pm 0.02$ \\
\hline

ATLAS & $B^0_s \rightarrow J/\psi \phi$ & 10 GeV & $52.75 \pm 0.28$ & $60 \pm 140$ & $0.53 \pm 0.09$ \\
ATLAS & $B^0_s \rightarrow J/\psi \phi$ & 15 GeV & $43.46 \pm 0.24$ & $-30 \pm 150$ & $0.60 \pm 0.10$ \\
\hline
\hline
        \end{tabular}
     \caption{\label{tab:table1}Number of reconstructed
$\rm{B^0_s}$ and $\rm{X^{\pm}(5568)}$ candidate events and the relative reconstruction efficiency $\rm{\epsilon_X}$ 
for the search of the $\rm{X^{\pm}(5568) \rightarrow B^0_s\pi^\pm}$ decay at D0, LHCb, CDF, CMS and ATLAS experiments~\cite{ref:D01}, \cite{ref:D02}, 
\cite{ref:LHCb}, \cite{ref:CDF}, \cite{ref:CMS}, \cite{ref:ATLAS}.
}
  \end{center}
\end{table*}

\section{Method for the statistical combination}
\label{Sec2}

The statistical combination of the upper limits on $\rm{\rho_X}$ uses the asymptotic frequentist $\rm{CL_s}$ method~\cite{ref:CLs1}\cite{ref:CLs2}. The likelihood used for the limit extraction is:

\begin{eqnarray}\nonumber
\mathcal{L} &=& \rm{\Pi_{i} \Pi_{j}}  \,
    \rm{\mathcal{G}[N_{ij}^{obs}(B^0_s)|N_{ij}^{exp}(B^0_s),\sigma(N_{ij}(B^0_s))]} \times \\
 && \rm{\mathcal{G}[\epsilon_{ij}^{obs}(X)|\epsilon_{ij}^{exp}(X),\sigma(\epsilon_{ij}(X))]} \times \\ \nonumber
 && \rm{\mathcal{G}[N_{ij}^{obs}(X)|N_{ij}^{exp}(B^0_s) \cdot \rho_X \cdot \epsilon_{ij}^{exp}(X),\sigma_{ij}(N_{X})]} 
\label{eq:eq2}
\end{eqnarray}


\noindent where $\rm{N_{ij}^{obs}}$ is the number of observed ($\rm{B^0_s}$ or X) events, 
$\rm{\sigma_{ij}(N)}$ is the measured uncertainty on N,
$\rm{N_{ij}^{exp}}$ is the number of expected ($\rm{B^0_s}$ or X) events,
$\rm{\epsilon^{ij}_X}$ is the relative reconstruction efficiency and $\rm{\mathcal{G}[x|\mu,\sigma(x)]}$ indicates a Gaussian with mean 
$\mu$ and sigma $\sigma$. 
The product indexes {\it{i}} and {\it{j}} run over the experiments and the analysis channels, 
respectively: $\rm{i=\{ATLAS,CDF, CMS, LHCb\} }$ and $\rm{j=\{hadronic, semileptonic \} }$.
The measurements of different experiments are assumed to be uncorrelated.

The limit extraction uses the {\it{RooStats}} package~\cite{ref:Roostats}. The variables with the ${^{obs}}$ suffix in Equation~2 are the observables, the
variables with the ${^{exp}}$ suffix are nuisance parameters and the $\rm{\sigma}$ are taken as constants.

In order to test the consistency of this approach with the results obtained by the experiments, the upper limit at 95\% confidence level has
been extracted for each experiment separately, using the input values reported in Table~\ref{tab:table1}.
Systematic uncertainties are added in quadrature to the statistical ones. Inputs values for systematic uncertainties are taken by the respective papers, 
except for CMS where we referred to~\cite{ref:CMSthesis}, as they were not reported in~\cite{ref:CMS}.

\begin{figure}[ht]
\centering
  \includegraphics[width=0.45\textwidth]{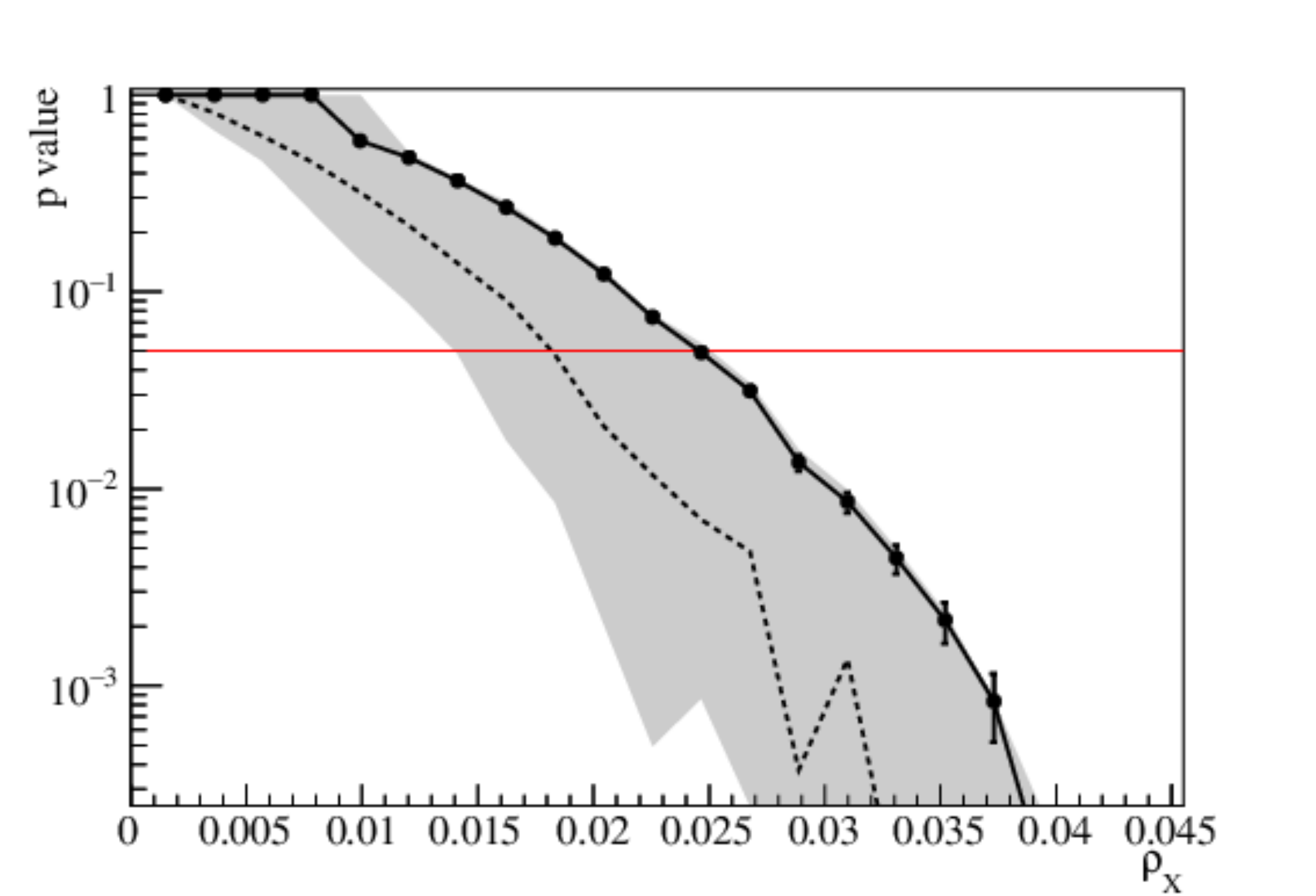}
  \caption{p-value as function of the parameter of interest $\rm{\rho_X}$ for LHCb 10 GeV bin. Black points are the computed
    observed CLs values for each value of $\rho_{X}$. The observed exclusion value at 95\% confidence level is given by the value
    of $\rm{\rho_X}$ corresponding to a p-value of 0.05 (indicated by the horizontal red line). The plot also shows the expected median
    value (dashed line) and the $\pm $1 $\sigma$ band (gray area).}
\label{fig:figure2}
\end{figure}

As an example, Fig.~\ref{fig:figure2} shows the results obtained running 10k pseudo-experiments for each of the 20 points of $\rm{\rho_X}$, 
for the $\rm{p_T(B^0_s)>}$ 10 GeV bin 
with LHCb inputs. The plot reports the observed CLs for each point as well  as the expected CLs and the $\rm{\pm 1 \sigma}$ band. 
The exclusion limit at 95\% CL is given by the interpolated value of $\rm{\rho_X}$ for $\rm{CLs^{obs}=0.05}$. 
The resulting limit is $\rm{\rho_X^{obs | LHCb}(p_{T}>~10GeV)}<$2.45\% at 95\% CL with an uncertainty of $\rm{\pm 0.04\%}$ from the numerical procedure of the limit extraction. 
The result is fully consistent with the corresponding limit reported by LHCb: $\rm{\rho_X}<$2.4\% at 95\% CL.

The same procedure has been applied to each $\rm{p_T}$ bin of each experiment, always finding good agreement with the results published by the four Collaborations.
The comparison is summarised in Table~\ref{tab:table2}. 

\begin{table*}[h!]
  \begin{center}
        \begin{tabular}{c|c|c|c}

\textbf{Experiment} & \textbf{$p_T$ cut} & \textbf{Experimental limit (\%)} & \textbf{Limit obtained in this work (\%)}  \\
\hline
\hline

ATLAS & 10 GeV & 1.5 & 1.48 $\pm$ 0.04 \\
ATLAS & 15 GeV & 1.6 & 1.70 $\pm$ 0.07 \\
\hline

CDF & 10 GeV & 6.7 & 6.69 $\pm$ 0.12  \\
\hline

CMS & 10 GeV & 1.1 & 1.15 $\pm$ 0.04  \\
CMS & 15 GeV & 1.0 & 1.04 $\pm$ 0.05  \\
\hline

LHCb & 10 GeV & 2.4 & 2.45 $\pm$ 0.04 \\
LHCb & 15 GeV & 2.0 & 2.19 $\pm$ 0.05 \\
\hline
\hline
        \end{tabular}
     \caption{\label{tab:table2} Comparison between the upper limit on $\rho_X$ as measured by the ATLAS, CDF, CMS and LHCb Collaborations and the reproduction of the results of this work. 
For ATLAS, CMS and LHCb the limits are extracted for each bin of p$_T(\rm{B^0_s})$.}
  \end{center}
\end{table*}

\section{Results}
\label{Sec3}

The limit for the combination is extracted by repeating the procedure and using the same inputs from Table~\ref{tab:table1}, separately for $\rm{p_T>}$ 10 and  
15 GeV (as LHCb is the only experiment having done the analysis for $\rm{p_T>}$ 5 GeV, no combination is performed in this case).

\begin{figure}[ht]
\centering
  \includegraphics[width=0.45\textwidth]{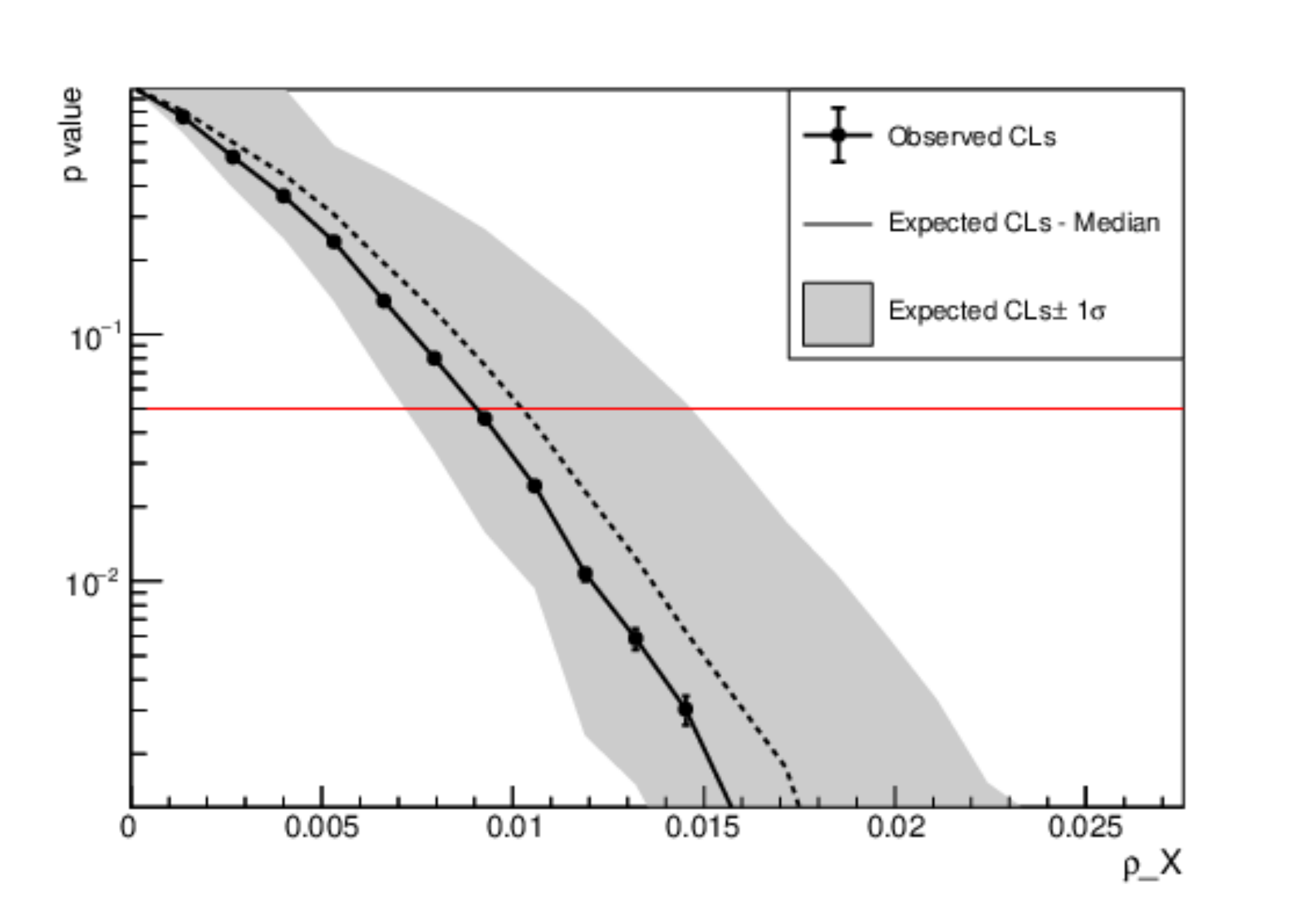}        
  \caption{p-value as function of $\rm{\rho_X}$ for the combination of the three LHC experiments for $\rm{p_T(B^0_s)>}$ 15 GeV.
    Black points are the computed observed CLs values for each value of $\rm{\rho_{X}}$. The observed exclusion value at 95\% confidence level is given by the value
    of $\rm{\rho_X}$ corresponding to a p-value of 0.05 (indicated by the horizontal red line). The plot also shows the expected median
    value (dashed line) and the $\pm $1 $\sigma$ band (gray area).}
\label{fig:figure3}
\end{figure}

Figure~\ref{fig:figure3} shows the CLs extraction for the combination of the LHCb, CMS and ATLAS results for $\rm{p_T(B^0_s \pi)> 15}$ GeV, with 50k pseudo-experiments per point. 

The combination gives the following upper limits at 95\% C.L.: 
\begin{eqnarray}\nonumber
\rm{\rho_X^{ATLAS+CMS+LHCb}(p_{T}(B^0_s)>10 GeV) < 0.92\%} \nonumber \\ 
\rm{\rho_X^{ATLAS+CMS+LHCb}(p_{T}(B^0_s)>15 GeV) < 0.91\%} \nonumber
\label{eq:eq3}
\end{eqnarray}

The combination of the LHC results improves the upper limit on $\rm{\rho_X}$
respect to the current best limits ($\rm{\rho_X < 1.1\% }$  for $\rm{p_T(B^0_s)>}$ 10 GeV and $\rm{\rho_X < 1.0\% }$  for $\rm{p_T(B^0_s)>}$ 15 GeV from CMS),
bringing both of them below the 1$\%$ level.
This result has to be compared with the D0 measurement of $(8.6\pm1.9(\rm{stat})\pm1.4(\rm{syst}))\%$ for the hadronic channel and
$(7.3^{+2.8}_{-2.4}(\rm{stat})^{+0.6}_{-1.7}(\rm{syst})\%$ for the leptonic channel.

The inclusion of the CDF result gives for the $\rm{p_T(B^0_s)>10}$ GeV case:
\begin{eqnarray}\nonumber
\rm{\rho_X^{ATLAS+CDF+CMS+LHCb}(p_{T}(B^0_s)>10 GeV) < 0.96\%} \nonumber \\ \nonumber
\label{eq:eq4}
\end{eqnarray}
The combined LHC+CDF limit is looser than the LHC-only one as CDF measures a number of $\rm{X^{\pm}(5568)}$ candidate events that is compatible with a $\sim1\sigma$ signal: $36\pm33$ (Table~\ref{tab:table2}).

The result from D0 is not included in the combination as D0 made a measurement of $\rm{\rho_X}$ and did not set an upper limit.
Moreover, some selection cuts in the D0 analysis are different from the other experiments. In particular the cut on
the transverse momentum of the $\rm{B^0_s \pi^{\pm}}$ pair, while all the other experiments have performed homogeneous analyses with the cut
on the transverse momentum of the $\rm{B^0_s}$, defining the analysis bin.
The described statistical procedure can be, however, applied to the D0 result alone. The combination of the hadronic and semi-leptonic channels
gives an upper limit on $\rm{\rho_X}$ of 12$\%$ at 95$\%$ CL.

\section{Conclusions}
\label{Sec4}

In conclusion, we have performed a statistical combination of the results from the ATLAS, CDF, CMS and LHCb experiments of the searches for the $\rm{X^\pm(5568)}$
state decaying into a $\rm{B^0_s\pi^\pm}$ pair.
The combination allows to set upper limits on the relative production rate of the $\rm{X^{\pm}(5568)}$ and $\rm{B^0_s}$ states, times the branching ratio for the
decay $\rm{X^{\pm}(5568) \rightarrow B^0_s\pi^\pm}$, $\rm{\rho_X}$, for the first time below 1$\%$ at 95\% CL. The most stringent limits are obtained when combining the results from the three LHC experiments: 
$\rm{\rho_X^{LHC} < 0.92\%}$ and $\rm{<0.91\%}$ at 95\% CL for $\rm{p_T(B^0_s)>}$ 10 and 15 GeV, respectively.

\section*{Acknowledgements}
\label{Ack}

The author is grateful to Dr. Richard Hawkings (CERN) for the review of the draft and the useful discussions.





\end{document}